# IDENTIFICATION OF THE OSCILLATION PERIOD OF CHEMICAL REACTORS BY CHAOTIC SAMPLING OF THE CONVERSION DEGREE

Marcin Lawnik, Marek Berezowski*

Silesian University of Technology, Faculty of Applied Mathematics, ul. Kaszubska 23, 44-100 Gliwice, Poland

To stabilise the periodic operation of a chemical reactor the oscillation period should be determined precisely in real time. The method discussed in the paper is based on adaptive sampling of the state variable with the use of chaotic mapping *to itself*. It enables precise determination of the oscillation period in real time and could be used for a proper control system, that can successfully control the process of chemical reaction and maintain the oscillation period at a set level. The method was applied to a tank reactor and tubular reactor with recycle.

## 1. INTRODUCTION

Stable state is a natural operational state of a chemical reactor (Berezowski, 1990; Gawdzik and Berezowski, 1987; Kubiček et al., 1980; Tsotsis and Schmitz, 1979). However, it may happen that the oscillation state is more advantageous for some reasons concerning the process, for example, due to a higher degree of conversion (Berezowski, 2011), involving the problem of automatic control of the reactor to stabilise the oscillation period (Antoniades and Christofides, 2001; Douglas, 1972; Luss and Amundson, 1967). In such a case, it is essential to identify the state variables in real time. The paper is focused on the method of identifying these variables by means of their proper sampling. The basic way of sampling was presented by (Shannon, 1949), where the reconstruction of a signal is achieved on the basis of evenly collected samples. In accordance with Nyquist–Shannon theorem, a signal should be sampled with the frequency that is at least twice as big as the boundary frequency of its spectrum. However, the determination of the boundary frequency is not always possible in practice. Under such circumstances, the sampling should be performed at uneven time intervals, for example: by adaptive sampling, in which the actual sampling moment depends on the previous sample value. In (Petkovski et al., 2006) an adaptive sampling algorithm based on Haar wavelet was presented. In turn, in (Feizi et al. 2010), the moment of sampling is determined by the previous sample value transformed by a specific function.

The scope of this paper is the adaptive sampling approach that is much easier than the so far described approaches. It is based on uneven adaptive sampling with the use of the chaotic representation of the tested signal (Berezowski and Lawnik, 2014) and does not require the use of additional functions or measurements in the time prescriptive points. The presented method works automatically.

---

*Corresponding author, e-mail: marek.berezowski@polsl.pl   

## 2. THE METHOD

The method is based on the representation of the sampled variable *to itself* in accordance with the following equations:

$$x_{k+1} = f(x_k) \tag{1}$$

$$t_{k+1} = \Omega(x_k)T + kT \tag{2}$$

where: $x$ is the state variable (for example: conversion degree), $T$ is the oscillation period, $t$ is the sampling moment, $k$ is the sample number and $\Omega(x_k)$ - normalisation coefficient. Accordingly, for the initial value of $x_0$ the first sample should be collected at the moment $t_1 = \Omega(x_0)T$. The value of the sample is $x_1$. The second sample should be collected at the moment $t_2 = \Omega(x_1)T + T$ and its value is $x_2$, and so forth. Thus, the sampling moments are determined by the values of the variable. If Eq. (1) is a chaotic representation, the sampling described above reconstructs the graph of the variable during the entire oscillation period. If the value of $T$ is selected incorrectly, the graph derived from the sampling does not reconstruct the graph of the variable, and its points are dispersed on the plane. If coefficient $\Omega(x_k)$ is incorrectly selected, the graph of the variable is reconstructed, but in a range different than one oscillation period. To determine the oscillation period, $\Omega(x_k)$ needs not to be precisely selected. What matters is the precise determination of $T$. When $T$ is chosen incorrectly, the reconstructed graph becomes ambiguous (the figure shows a cloud of points – see Fig. 2 and Fig. 4). In the case when $\alpha$ denotes the conversion degree, it can be assumed that $\alpha_{min} = 0$, $\alpha_{max} = 1$, which gives $\Omega(\alpha_k) = \alpha_k$. In an example presented in the paper, these values were read from the graph simulation. In the case when the trajectory of the studied variable is complex, accurate computation of the period of oscillations in real time is virtually impossible.

Accordingly, the proposed method enables precise determination of the actual oscillation period and automatic control of the reactor. A mathematical model of a reactor was used, not a real one. The samples obtained from the following examples were not collected from measurements but from computer simulations.

## 3. TANK REACTOR

The dimensionless balance equations for a non-adiabatic tank reactor with ideal mixing are expressed as:
- mass balance:

$$\frac{d\alpha}{d\tau} + \alpha = \phi(\alpha, \Theta) \tag{3}$$

- heat balance:

$$\frac{d\Theta}{d\tau} + \Theta = \phi(\alpha, \Theta) + \delta(\Theta_H - \Theta) \tag{4}$$

- kinetic reaction function:

$$\phi(\alpha, \Theta) = Da(1-\alpha)^n \exp\left(\gamma \frac{\beta\Theta}{1+\beta\Theta}\right) \tag{5}$$

After setting the following values of the parameters: $Da = 0.15$, $\gamma = 15$, $n = 1.5$, $\beta = 2$, $\delta = 3$, $\Theta_H = -0.01$ the above model gives a periodic solution. To reconstruct this solution the degree of conversion should be sampled at demensionless time intervals $\tau_{k+1}$ determined by Equation (2), assuming that

$$\Omega(\alpha_k) = \frac{\alpha_k - \alpha_{min}}{\alpha_{max} - \alpha_{min}} \qquad (6)$$

Assuming that $\alpha_{min}$ = 0.607388, $\alpha_{max}$ = 0.954754 and $T$ = 1.81775965 the graph presented in Fig. 1. was plotted. For any other value of $T$ the sampling presents a cloud of points – see Fig. 2. Thus, the set value of $T$ is, at the same time, the real oscillation period.

In the course of a real process, a change in the value of $T$ may occur, for example, due to changes of the values of the reactor parameters. In such a case, the sampling will also result in a cloud of points, which may evoke a proper reaction of the control system.

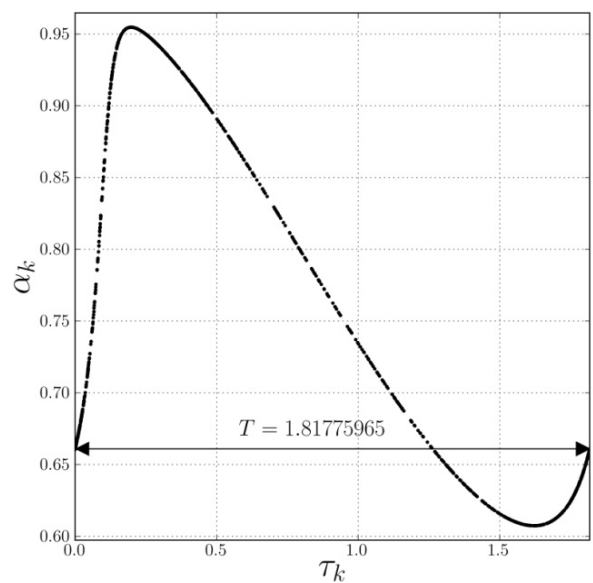

Fig. 1. Correct reconstruction of the conversion degree graph for the tank reactor; $T$ = 1.81775965

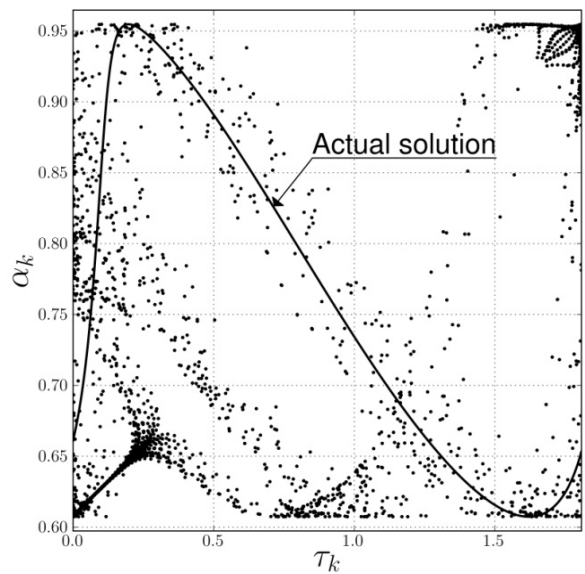

Fig. 2. Erroneous reconstruction of the conversion degree graph for the tank reactor (black dots); $T$ = 1.80775965

## 4. TUBULAR REACTOR WITH RECYCLE

The balance equations for a homogeneous non- adiabatic tubular reactor with recycle as expressed by (Berezowski, 2006; Berezowski, 2013; Reilly and Schmitz, 1966; Reilly and Schmitz, 1967):

- mass balance:

$$\frac{\partial \alpha}{\partial \tau} + \frac{\partial \alpha}{\partial \xi} = (1-f)\phi(\alpha, \Theta) \tag{7}$$

- heat balance:

$$\frac{\partial \Theta}{\partial \tau} + \frac{\partial \Theta}{\partial \xi} = (1-f)\phi(\alpha, \Theta) + \delta(\Theta_H - \Theta) \tag{8}$$

- kinetic reaction function:

$$\phi(\alpha, \Theta) = Da(1-\alpha)^n \exp\left(\gamma \frac{\beta \Theta}{1+\beta \Theta}\right) \tag{9}$$

- boundary conditions:

$$\alpha(\tau,0) = f\alpha(\tau,1); \quad \Theta(\tau,0) = f\Theta(\tau,1). \tag{10}$$

After setting the following values of the parameters: $Da = 0.15$, $\gamma = 15$, $n = 1.5$, $\beta = 2$, $\delta = 3$, $\Theta_H = -0.02157$, $f = 0.5$ the above model gives the solution of a discrete eight-periodic orbit. As the particular sampling moments are, in such case, integer values, recurrence Equation (2) should be transformed into:

$$L_{k+1} = Integer\{T\Omega_N[\alpha_k(\tau,1)]\} + kT \tag{11}$$

where *L* and *T* are integer numbers, expressing the sampling moment and oscillation period, respectively.

Unlike the continuous model, in the case presented in the paper, the periodic solution is a finite number of points. For the set parameter values, there are 8 points (eight-periodic orbit). This means that the recurrence algorithm (1)-(2) cannot generate a chaotic sequence and, consequently, discrete time graph cannot be reconstructed. Under such circumstances, an additional transformation that provides a chaotic solution should be introduced into the calculation procedure. For this purpose the normalisation coefficient $\Omega_N(\alpha_k)$ derived from Equation (11) is used, the value of which shall be derived from the *skew tent map*, expressed as the following recursive procedure:

$$\Omega_0 = \alpha_k \tag{12}$$

$$\Omega_{j+1} = \frac{\Omega_j}{p} \quad \text{for} \ \ 0 \leq \Omega_j \leq p \tag{13}$$

$$\Omega_{j+1} = \frac{1-\Omega_j}{1-p} \quad \text{for} \ \ p < \Omega_j \leq 1 \tag{14}$$

where *j* changes from 0 to *N*. This transformation provides a chaotic solution for any value of parameter *p* within the range 0 < *p* < 1.

Therefore, for an initial value of $\alpha_0$, the first sample should be collected at the moment $L_1 = Integer[T\Omega_N[\alpha_0(\tau,1)]]$. The value of the sample is $\alpha_1$. The second sample should be collected at the moment $L_2 = Integer[T\Omega_N[\alpha_1(\tau,1)]] + T$ and its value is $\alpha_2$, and so forth. Assuming that *p* = 0.3, $N = 100$ and $T = 8$ a discrete time graph of the reactor was reconstructed - see Fig. 3. For an erroneous assumption of the value of the oscillation period, for example: $T = 9$, an ambivalent solution was derived - see Fig. 4.

Just like in the case of the tank reactor, the discussed sampling method may be useful for controlling the operation of the tubular reactor with recycle (Antoniades and Christofides, 2001).

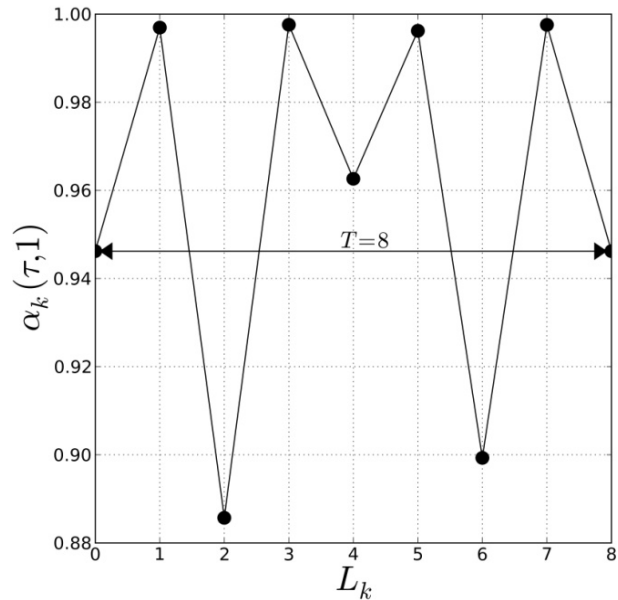

Fig. 3. Correct reconstruction of the conversion degree graph for the tubular reactor with recycle; $T = 8$

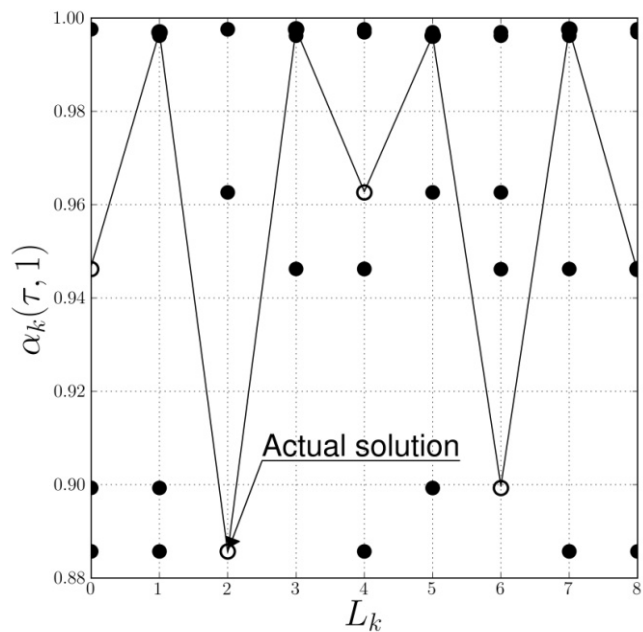

Fig. 4. Erroneous reconstruction of the conversion degree graph for the tubular reactor with recycle (black dots); $T = 9$

## 5. SUMMARY

The method of identifying the conversion degree in chemical reactors under the oscillation mode of operations was presented. In particular, the method enables precise determination of the oscillation period in real time. The discussed procedure is based on a chaotic representation of the state variable *to itself.* Accordingly, a proper control system can successfully control the chemical reaction process and maintain the oscillation period at a set level. The method was presented for a tank reactor and a tubular reactor with recycle. The corresponding graphs illustrate the influence of oscillation period value. If the value is selected incorrectly, the graph is ambivalent, rendering a cloud of points.

## SYMBOLS

| | |
|---|---|
| $c_p$ | heat capacity, kJ/(kg K) |
| $C_A$ | concentration of component *A*, kmol/m$^3$ |
| $Da$ | Damköhler number $\left（=\dfrac{V_R(-r_0)}{\dot{F}C_{A0}}\right)$ |
| $E$ | activation energy, kJ/kmol |
| $\dot{F}$ | volumetric flow rate, m$^3$ s |
| $(-\Delta H)$ | heat of reaction, kJ/kmol |
| $K$ | reaction rate constant, $1/(s \cdot (m^3 \text{ kmol}^{-1})^{n-1})$ |
| $L$ | length, m |
| $n$ | order of reaction |
| $(-r)$ | rate of reaction, $(= KC^n)$, kmol/m$^3$ |
| $R$ | gas constant, kJ/(kmol K) |
| $t$ | time, s |
| $T$ | period |
| $V$ | volume, m$^3$ |
| $z$ | position, m |

*Greek symbols*

| | |
|---|---|
| $\alpha$ | degree of conversion $\left(=\dfrac{C_{A0}-C_A}{C_{A0}}\right)$ |
| $\beta$ | dimensionless number related to adiabatic temperature increase $\left(=\dfrac{(-\Delta H)C_{A0}}{\Gamma_0 \rho c_p}\right)$ |
| $\gamma$ | dimensionless number related to activation energy $\left(=\dfrac{E}{R\Gamma_0}\right)$ |
| $\delta$ | dimensionless heat exchange coefficient $\left(=\dfrac{A_q k_q}{\rho c_p \dot{F}}\right)$ |
| $\Gamma$ | temperature, K |
| $\Theta$ | dimensionless temperature $\left(=\dfrac{\Gamma - \Gamma_0}{\beta \Gamma_0}\right)$ |
| $\xi$ | dimensionless position $\left(=\dfrac{z}{L}\right)$ |
| $\rho$ | density, kg/m$^3$ |

| | | |
|---|---|---|
| $\tau$ | dimensionless time $\left(=\dfrac{\dot{F}}{V_R}t\right)$ | |

*Subscripts*

| | |
|---|---|
| *max* | refers to the maximum of state variable |
| *min* | refers to the minimum of state variable |
| *0* | refers to the feed; refers to the initial state |
| *H* | refers to the temperature of cooling medium |
| *R* | refers to the reactor |